\pdfoutput=1
\documentclass[11pt]{article}

\usepackage[final]{acl}
\usepackage{times}
\usepackage{latexsym}

\usepackage[T1]{fontenc}
\usepackage[utf8]{inputenc}
\usepackage{microtype}
\usepackage{inconsolata}
\usepackage{graphicx}
\usepackage{tipa}
\usepackage{subfigure}
\usepackage{placeins}

\title{From perception to production: how acoustic invariance facilitates articulatory learning in a self-supervised vocal imitation model}

\author{
  \textbf{Marvin Lavechin\textsuperscript{1}},
  \textbf{Thomas Hueber\textsuperscript{2}}
  \\
  \textsuperscript{1}Computational Psycholinguistics lab, Massachusetts Institute of Technology, United States
  \\
  \textsuperscript{2}Univ. Grenoble Alpes, CNRS, Grenoble INP, GIPSA-lab, Grenoble, France
\\
  \small{
    \textbf{Correspondence:} \href{mailto:email@domain}{marvinlavechin@gmail.com}
  }
}

\begin{document}
\maketitle
\begin{abstract}
Human infants face a formidable challenge in speech acquisition: mapping extremely variable acoustic inputs into appropriate articulatory movements without explicit instruction. We present a computational model that addresses the acoustic-to-articulatory mapping problem through self-supervised learning. Our model comprises a feature extractor that transforms speech into latent representations, an inverse model that maps these representations to articulatory parameters, and a synthesizer that generates speech outputs. Experiments conducted in both single- and multi-speaker settings reveal that intermediate layers of a pre-trained wav2vec 2.0 model provide optimal representations for articulatory learning, significantly outperforming MFCC features. These representations enable our model to learn articulatory trajectories that correlate with human patterns, discriminate between places of articulation, and produce intelligible speech. Critical to successful articulatory learning are representations that balance phonetic discriminability with speaker invariance -- precisely the characteristics of self-supervised representation learning models. Our findings provide computational evidence consistent with developmental theories proposing that perceptual learning of phonetic categories guides articulatory development, offering insights into how infants might acquire speech production capabilities despite the complex mapping problem they face.
\end{abstract}

\section{Introduction}

Speech development undergoes dramatic changes during infancy, progressing from early non-linguistic vocalizations like crying and gurgling, to meaningful speech with adult-like intonation patterns \cite{stark1980stages}. One key mechanism driving this development is \textit{vocal imitation}, whereby infants attempt to match the vocalization patterns of their caregivers. \citet{kuhl1996infant} found that infants' vowel-like sounds become more acoustically distinct with age and that exposure to specific vowels can elicit matching vocalizations, suggesting that infants use auditory targets from speech input to guide their own vocal development. While evidence of imitation in prelinguistic infants is still debated, more evidence can be found in older children who have been found to actively imitate arbitrary sounds directed to them by adults \cite{ross2003generalized,jones2007imitation,pelaez2018infant}. These studies provide compelling evidence that infants use aspects of their auditory inputs as targets for their own vocalizations; however, they leave open a fundamental question: how do infants achieve this imitation when they cannot see the complex movements inside a speaker's vocal tract?

One central challenge in understanding early speech production learning is how infants map auditory targets onto motor actions. This problem is challenging due to three key factors: 1) the many-to-many mapping between sounds and articulatory gestures \cite{atal1978inversion}, where identical sounds can result from different movements and vice versa; 2) the normalization problem, where infants must extract invariant phonetic features from diverse speakers and map these onto their own articulatory system; and 3) the absence of explicit feedback on articulation correctness, requiring infants to rely on some form of unsupervised or self-supervised learning by comparing their vocalizations with heard targets to adjust their movements.

Computational models provide a way to explore how the acoustic-to-articulatory learning might unfold, enabling researchers to test hypotheses about representations and learning mechanisms that could support speech production learning without explicit feedback. A range of computational models has been proposed to investigate how vocal learners — whether infants or artificial agents — map auditory targets to motor commands. Early models often used supervised learning with predefined phonetic targets and explicit error signals to train articulatory controllers \cite{browman1992articulatory,guenther1995speech}. More recent approaches have shifted toward unsupervised or self-supervised learning \cite{rasilo2017online,beguvs2023articulation}, including neural networks that integrate reinforcement with self-organization \cite{warlaumont2013prespeech} or curiosity-driven learning \cite{moulin2014self}. Closer to our work is \citet{georges2024decode}, who introduce a self-supervised model that learns to control an artificial vocal tract through imitation. However, their approach has key limitations: the model is trained on hyper-articulated speech produced by a single speaker and fails to learn human-like articulatory trajectories.

In this work, we propose a self-supervised imitation learning model that addresses the acoustic-to-articulatory mapping problem. Our model learns to translate auditory input into articulatory gestures by minimizing the distance between input and imitated speech representations. We systematically compare how different auditory representations -- from basic acoustic features (MFCCs) to increasingly abstract representations from wav2vec 2.0 layers \cite{baevski2020wav2vec} -- affect articulatory learning. These representations act as computational analogs of increasingly abstract speech sound representations infants develop during their first year \cite{lavechin2024modeling,poli2024modeling}. 

With a single French speaker, our model learns articulatory trajectories that correlate with human patterns and effectively distinguish between places of articulation. Going further, we evaluate our model in a multi-speaker scenario, where we demonstrate that speech representations that optimally balance phonetic information with speaker invariance enable our model to generalize across speakers and successfully discriminate between places of articulation. 

Our findings suggest a developmental scenario where perceptual acoustic invariance is crucial for articulatory learning. The observed U-shaped performance curve across wav2vec 2.0 layers suggests an optimal abstraction level: lower layers fail to normalize across speakers, while higher layers lose critical phonetic distinctions needed to solve the acoustic-to-articulatory inverse problem.

\section{Methods}

\subsection{The imitation model}

We provide an overview of all of the components of our imitation model.

\begin{figure*}[h]
   \centering
 \includegraphics[width=.98\textwidth, trim={0 2.9cm 0 0}, clip]{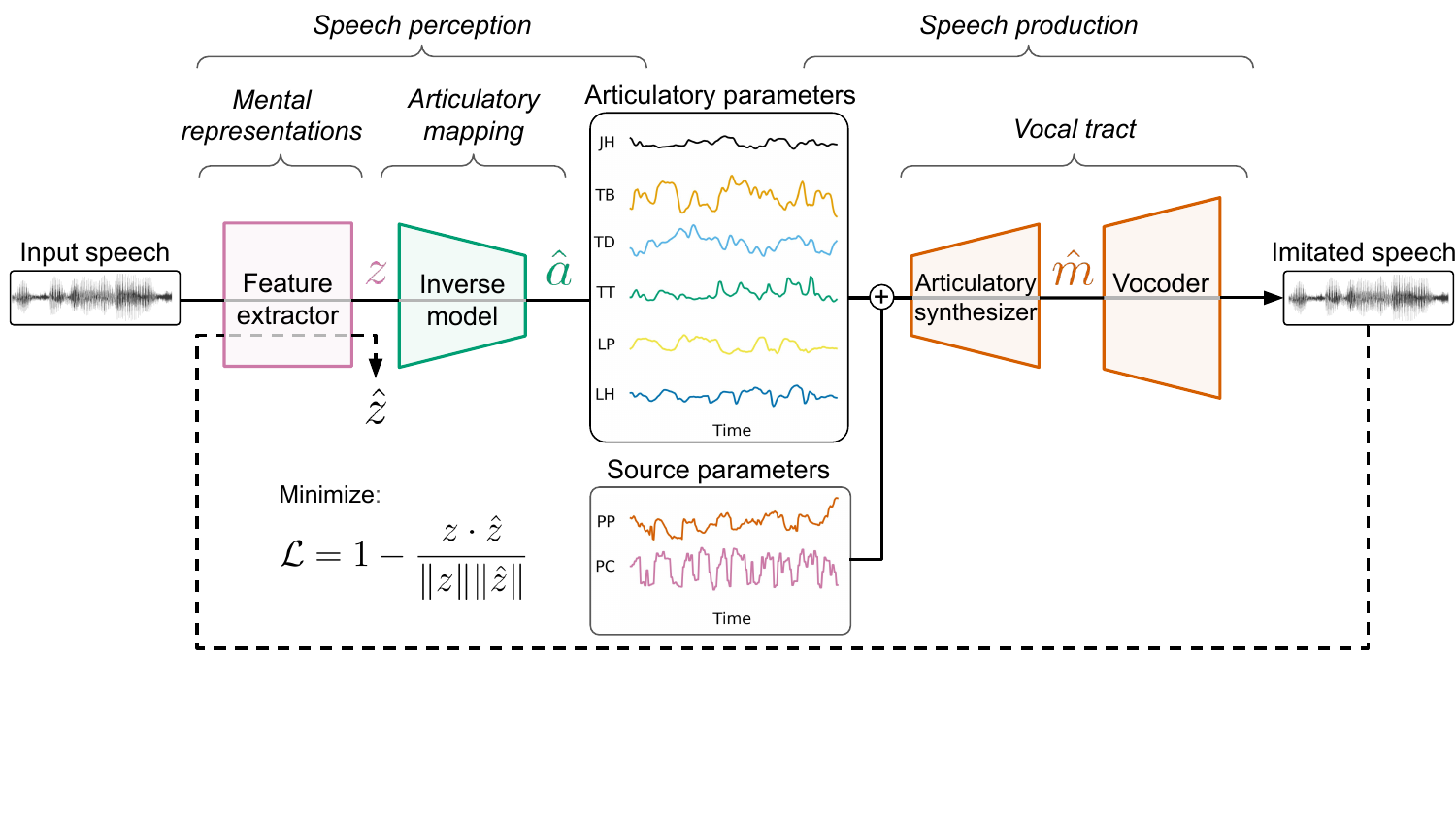} 
 \caption{Architecture of the speech imitation model. The system consists of four components: (1) a feature extractor that encodes the raw speech waveform into a sequence of representations, (2) an inverse model that maps these representations to articulatory parameters, (3) an articulatory synthesizer that combines these parameters with source parameters to generate mel-spectrograms, and (4) a vocoder that converts mel-spectrograms to waveforms. During training, only the inverse model is optimized to minimize a cosine similarity loss function between the input speech segment and the imitated speech segment in the representational space. All other components remain frozen, allowing targeted investigation of sensorimotor mapping acquisition. JH: Jaw Height, TB: Tongue Body, TD: Tongue Dorsum, TT: Tongue Tip, LP: Lip Protrusion, LH: Lip Height, PP: Pitch Period, PC: Pitch Coefficient.}
 \label{fig:model}
\end{figure*}

\noindent\textbf{Feature extractor.} The feature extractor transforms input speech signals into feature vectors $z_t$ for each frame $t$. In this work, we investigate two approaches: 1) a baseline extractor that derives low-level acoustic features through MFCCs, and 2) representations from the transformer layers of a pretrained wav2vec 2.0 model that capture higher-level properties of speech. These features serve as computational analogs to the mental encoding processes humans might employ when perceiving speech, creating internal representations that guide subsequent articulation. By comparing different feature extractors, we can determine which representations yield optimal articulatory trajectories.

\noindent\textbf{Inverse model.} The inverse model maps the extracted feature vectors $z_t$ to estimated articulatory parameters $\hat{a_t}$, simulating the human's brain ability to infer the physical configurations required for speech production. For each frame $t$, the inverse model predicts a comprehensive set of articulatory parameters including jaw height (JH), tongue body (TB), tongue dorsum (TD), tongue tip (TT), lip protusion (LP), and lip height (LH). This component is trained by minimizing a cosine distance loss function between the features extracted from the input speech segment $z_t$ and those from the imitated speech segment $\hat{z_t}$. Through this process, the inverse model learns to decode the input features into articulatory commands that can be executed by the subsequent production system.

\noindent\textbf{Articulatory synthesizer.} The articulatory synthesizer processes both the predicted articulatory parameters and two source parameters $s_t$ computed directly from the input speech: pitch period (PP) and pitch coefficient (PC), which represent fundamental frequency and harmonicity respectively. Source parameters characterize vocal fold vibration (the "voice" itself), while articulatory parameters describe the physical configuration of the vocal tract (jaw, tongue, lips) that shapes this voice. These source parameters are directly extracted from the input speech and passed through unchanged, rather than being predicted by the model, as our work focuses specifically on the acoustic-to-articulatory inverse mapping problem. From this combined input of articulatory $a_t$ and source parameters $s_t$, the synthesizer generates mel-spectrograms $\hat{m_t}$. The articulatory synthesizer serves as a computational analog to the human vocal tract, translating articulatory configurations into acoustic representations.

\noindent\textbf{Vocoder.} The vocoder converts mel-spectrograms into the final imitated speech segment, completing the production pathway. The vocoder effectively "gives voice" to the articulatory plans, producing acoustic output that mimics the perceived input stimuli. This final component enables objective and perceptual evaluation of imitation quality, measuring how successfully the model has captured and reproduced the target speech. 

\subsection{Training procedure}

We focus on the acoustic-to-articulatory mapping by optimizing only the inverse model while keeping all other components frozen. While this is a simplification of actual infant development -- who experience simultaneous changes in how they perceive speech sounds \cite{werker1984cross,kuhl1995linguistic} and in the physical structure of their vocal tracts \cite{Archibald1994PhonologicalD,sasaki1977postnatal} -- this experimental design allows us to rigorously analyze the development of inverse mappings from perception to articulation without confounding variables (but see \nameref{sec:limitations}).

\subsection{Datasets}

\noindent\textbf{PB2009.} This corpus consists of 37 minutes of French speech (from simple vowels and consonants to full sentences) from a single male speaker \cite{badin_2022_6390598}. The participant's tongue, jaw, and lips movements were recorded using a 2-D ElectroMagnetic Articulograph (EMA) with a sampling rate of 200 Hz. HMM-generated phonetic transcriptions which were manually corrected are provided. PB2009  was split into 64\% training, 16\% validation, and 20\% test sets.

We use this corpus to train the vocoder and the articulatory synthesizer (not central to this paper), and to train the complete imitation model in a single speaker setting. This dataset was crucial for establishing baseline performance under optimal conditions (same speaker for input and synthesizer) and for validating our learned articulatory trajectories against ground truth measurements.

\noindent\textbf{Barnig's text-to-speech (TTS).} This dataset of 20 hours of read speech in Luxembourgish, German, French, English, and Portuguese from 18 speakers was used to train the vocoder. 

\noindent\textbf{Audiocite.} We subsampled 104 hours of French read speech produced by 8 speakers (4 females) from the Audiocite corpus \cite{felice2024audiocite}. We used 100 hours in its raw, untranscribed form to train our imitation model. The remaining 4 hours were set aside as test audio files, which we automatically transcribed, aligned, and spliced to remove non-speech segments using Whisper-small \cite{radford2023robust,bain2023whisperx}. This setup allows us to assess the model's generalization capabilities across different vocal tract configurations and speaking styles.

\noindent\textbf{Librispeech.} We used Librispeech \cite{panayotov2015librispeech} exclusively for probing experiments to analyze the phonetic and speaker information encoded in different layers of wav2vec 2.0 via linear classification. We used the original training set to train the linear probes and the test set to compute phone and speaker accuracies. 

\subsection{Implementation details}

\noindent\textbf{The model.} The feature extractor uses either:  1) a baseline MFCC extractor providing 39-dimensional features (13 base coefficients plus first and second derivatives), implemented via \textit{torchaudio} for gradient backpropagation support \cite{yang2022torchaudio}, or 2) a wav2vec 2.0 model\footnote{\url{https://huggingface.co/facebook/wav2vec2-base-10k-voxpopuli}} with 12 layers and hidden size of 768 pretrained on VoxPopuli \cite{Wang2021VoxPopuliAL}. The inverse model employs a 2-layer bidirectional LSTM with 64-dimensional hidden states, outputting 6-dimensional articulatory parameter vectors. Each inverse model was trained using Adam's scheduler with a learning rate (lr) of $1.7 \times 10^{-3}$. The articulatory synthesizer uses 4 feed-forward (FF) layers (512 units each) and was pre-trained on PB2009 using Adam's scheduler ($lr = 5 \times 10^{-4}$, \citeauthor{badin_2022_6390598}, \citeyear{badin_2022_6390598}). The vocoder uses a HiFi-GAN architecture \cite{kong2020hifi} pre-trained on Barnig's TTS corpus \cite{mbarnig_lb_de_fr_en_pt_2022} and fine-tuned on PB2009 \cite{badin_2022_6390598} using Adam's scheduler ($lr = 2 \times 10^{-4}$). All components except the inverse model are kept frozen during imitation learning.

\noindent\textbf{Data processing.} All audio samples were preprocessed to 16 kHz single-channel recordings.

Regarding acoustic features, MFCCs and mel-spectrograms were computed using a short-time Fourier transform with a frame length of 640 samples and hop size of 320 samples, followed by a mel-filterbank projection with 80 mel bands spanning frequencies from 0 to 8 kHz. MFCCs were extracted as 13-coefficient representations, which were z-scored, with first and second derivatives. This gives us one feature vector for every 20 milliseconds of audio, which matches the frame rate used in wav2vec 2.0.

Regarding articulatory features, EMA trajectories from PB2009 were downsampled at 50 Hz to match the acoustic frame rate. Similarly to \citet{georges2024decode}, raw EMA coordinates were projected into lower dimension vectors, referred as articulatory parameters (depicted in Figure \ref{fig:model}), using guided Principal Component Analysis (PCA) \cite{maeda1990compensatory, Serrurier2012}. Guided PCA begins by extracting a specific articulatory parameter (e.g., Jaw Height) from a designated set of measurements (e.g., lower incisor coil data). The contribution of this parameter to other articulator movements is then estimated using linear regression. This contribution is subtracted from the original measurements of other articulators, creating "residual movements" that are independent of the first parameter. PCA is then applied to these residual movements to extract the next articulatory parameter. This sequential process of extraction, regression, subtraction, and further PCA is repeated until all relevant articulatory parameters (JH, TB, TD, TT, LP, LH) are derived \cite{maeda1990compensatory}.

\subsection{Evaluation metrics}

We evaluate our model using complementary metrics that assess both the learned articulatory trajectories and the resulting imitated speech.

\noindent\textbf{Correlation with ground-truth articulatory parameters.} Correlations were computed as follows: For each articulatory parameter (JH, TB, TD, TT, LP, LH), we computed Pearson's correlation coefficient between the predicted and ground-truth trajectories across all time frames in the test set. The final correlation score represents the average correlation across all six articulatory parameters. This analysis enables us to evaluate whether our model learns physiologically realistic motor patterns rather than solutions that are acoustically plausible but physically implausible from an articulatory standpoint.

\noindent\textbf{Place of articulation ABX score.} Since human articulatory trajectories naturally organize consonants by place of articulation, we evaluate whether our learned articulatory trajectories distinguish between different places of articulation. 

We do so using the ABX discrimination test \cite{schatz2013evaluating}. We extract vowel-consonant-vowel sequences and construct triplets where A and X are instances of the same consonant (potentially in different vocalic contexts), while B has a different place of articulation but the same manner. The model is considered correct if $d(A,X) < d(B,X)$, with $d$ computed using cosine distance along the shortest dynamic time warping path -- see \citet{georges2024decode} for implementation details.

\noindent\textbf{Phone and speaker accuracy.} We separately evaluate wav2vec 2.0's layers by training linear probes that measure: (1) Phone accuracy using a linear CTC model with 256-dimensional projection trained on LibriSpeech phonemic transcriptions. Phone accuracy is calculated as (1 - PER), where PER (Phone Error Rate) is computed using frame-level phoneme predictions, and phonemic labels are derived from orthographic transcriptions using the LibriSpeech lexicon with grapheme-to-phoneme (G2P) conversion; (2) Speaker identification accuracy using a linear classifier over frame-level features for predicting speaker identity at each audio frame.
 
\noindent\textbf{Intelligibility of the imitated speech.} To evaluate how well our model preserves linguistic content in its imitations, we measure the intelligibility of the synthesized speech output. For the 8 French speakers of Audiocite, we process test utterances through our imitation model and transcribe the resulting imitated speech using Whisper-small \cite{radford2023robust}. We then compute WER by comparing: (1) the reference transcription of the original input speech, and (2) the transcription of our model's imitated speech. This evaluation presents a more substantial cross-speaker generalization challenge, as our model must map diverse acoustic inputs to articulatory patterns appropriate for a vocal tract with different characteristics than those of the original speakers. 

\section{Experiments}

Audio examples of our model's imitation across different experimental conditions are available at \url{https://marvinlvn.github.io/projects/from_perception_to_production}.

\subsection{Single-speaker setting}

Here, we focus on evaluating how well the model succeeds in learning articulatory trajectories ($\hat{a}$ in Figure \ref{fig:model}) that resemble those of a human speaker.

\subsubsection{Correlation with ground truth articulatory trajectories}

\begin{figure}[htbp]
    \centering
    \includegraphics[width=\columnwidth]{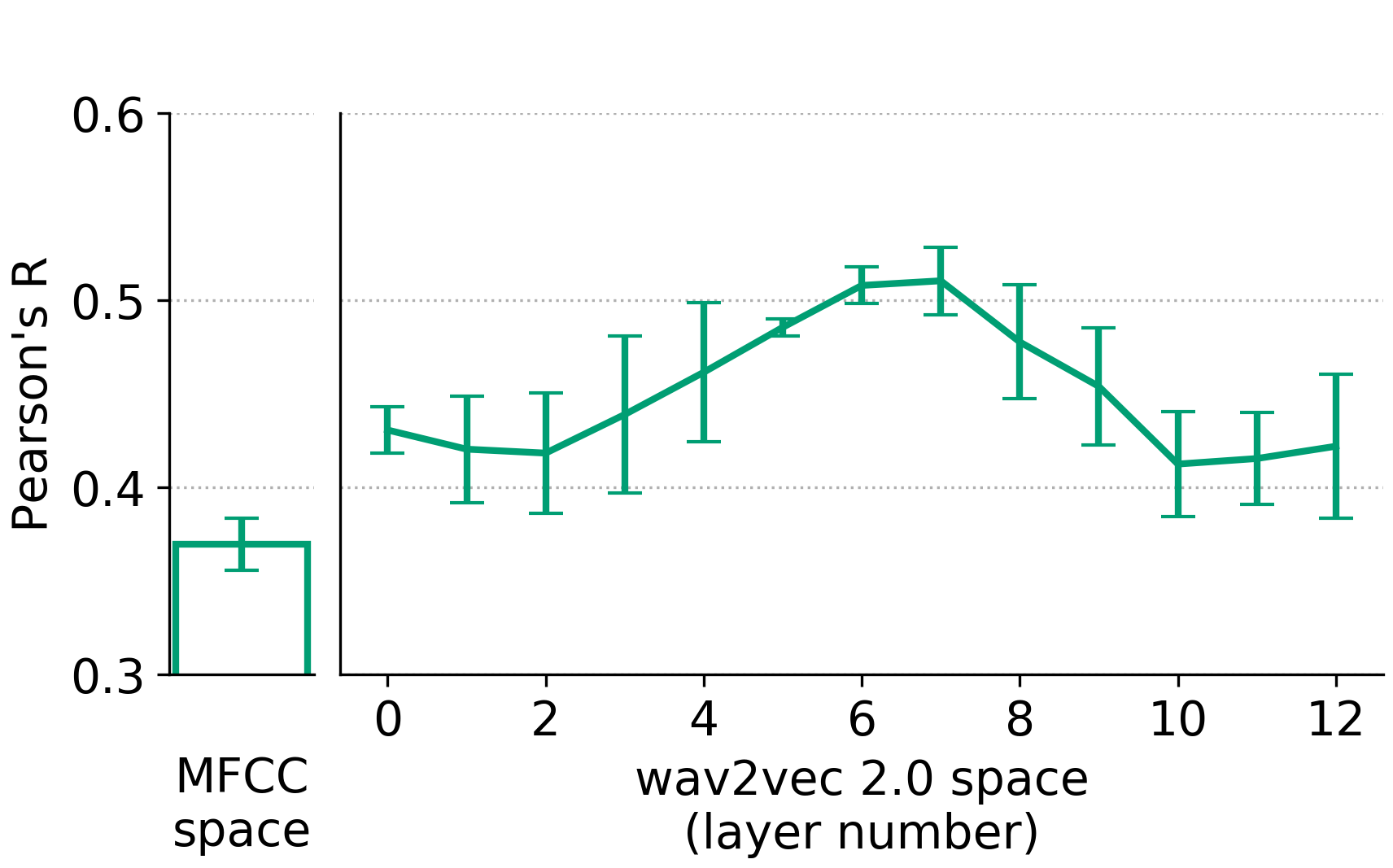}
    \caption{Pearson's $R$ correlation between predicted and ground truth articulatory parameters for the model imitating in the MFCC feature space (left) versus the wav2vec 2.0 space at different layers (right). Error bars represent standard deviation obtained across 5 different splits of the PB2009 dataset.}
    \label{fig:pb2009_pearson}
\end{figure}

Figure \ref{fig:pb2009_pearson} shows correlations between predicted and ground-truth articulatory trajectories on PB2009. When using MFCC features to encode speech input, the resulting articulatory predictions showed a moderate correlation with ground-truth parameters ($R \simeq 0.37$). In contrast, wav2vec 2.0 embeddings as speech representations yielded substantially stronger correlations in an inverted U-shaped pattern across layers. Peak performance occurred at intermediate layers (6-7), where articulatory predictions reached correlations of approximately $0.51$ with ground truth measurements -- a 38\% improvement over the MFCC baseline. Both lower layers (0-2, $R \simeq 0.42$) and higher layers (10-12) performed worse, suggesting articulatory-relevant information is lost at both extremes. Audio examples are available at \url{https://your-website.com}

These findings show our model learns human-like articulatory trajectories without explicit supervision, with the representational space used to optimize the loss significantly affecting trajectory quality (see Appendix \ref{sec:appendixA} for examples of trajectories). 

\subsubsection{Organization along the place of articulation (single-speaker)}

\begin{figure}[htbp]
    \centering
    \includegraphics[width=\columnwidth]{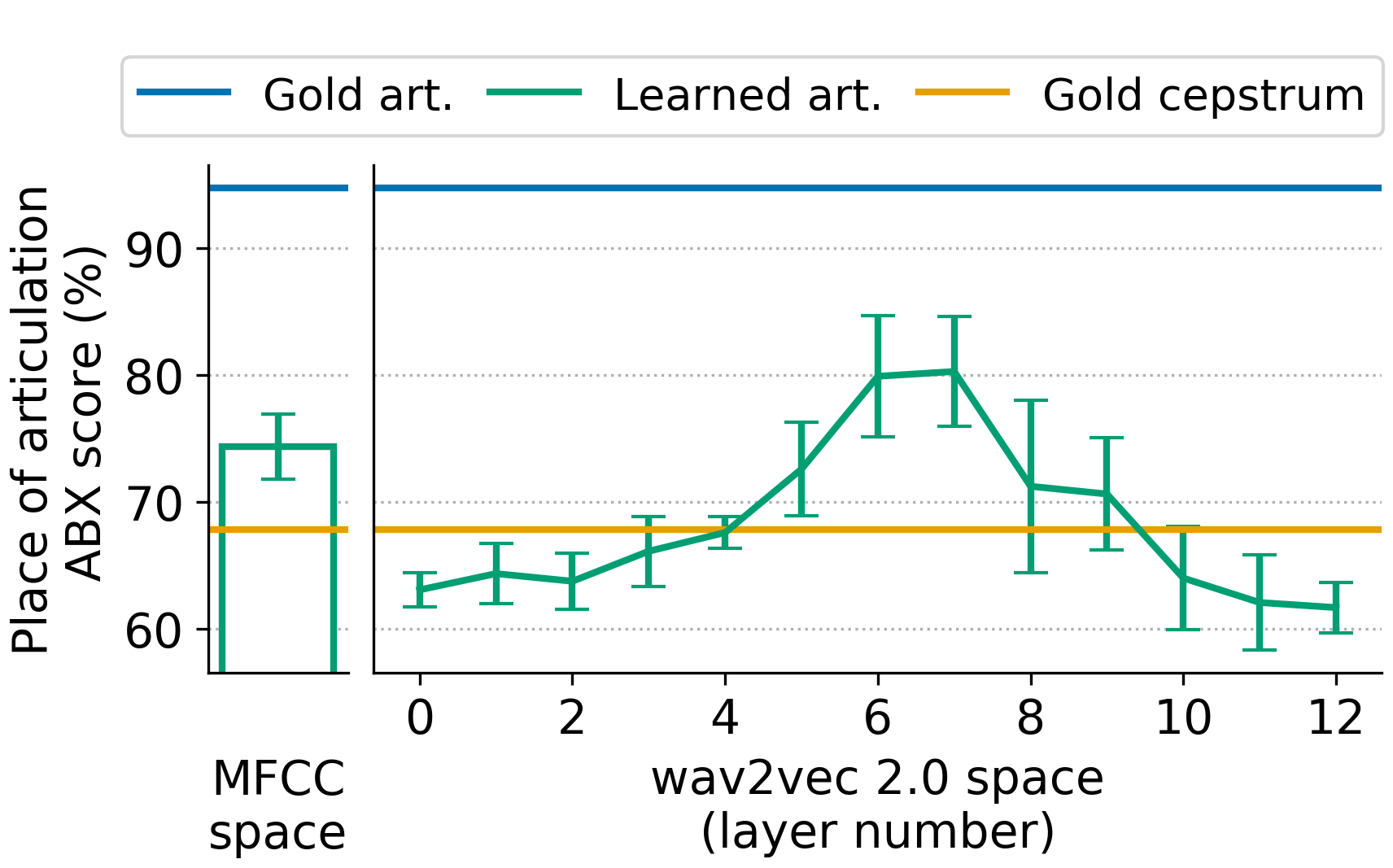}
    \caption{Place of articulation ABX scores (\%) obtained by the articulatory parameters learned by the model imitating in the MFCC feature space (left) versus the wav2vec 2.0 space (right). Horizontal lines show scores obtained by ground truth articulatory parameters (blue) and MFCCs extracted directly from the input audio (yellow). Error bars represent standard deviation obtained across 5 different splits of the PB2009 dataset.}
    \label{fig:pb2009_abx_place}
\end{figure}

Human articulation organizes consonants by place (labial, coronal, dorsal), with distinct configurations for each. Here, we measure the extent to which our model captures these distinctions.

Figure \ref{fig:pb2009_abx_place} shows the place of articulation ABX discrimination score for articulatory parameters learned by our model across different feature spaces. In the MFCC feature space, the inverse model learned articulatory trajectories that achieved an ABX score of approximately 74\%. This exceeds the acoustic baseline (orange horizontal line) of 68\% obtained from using MFCCs directly for discrimination, indicating that our model transforms these basic acoustic features into articulatory patterns that better distinguish place of articulation. However, this remains well below the ground-truth articulatory topline (blue horizontal line) of approximately 94\%, which represents optimal discrimination based on human trajectories.

In the wav2vec 2.0 feature space, we observe another inverted U-shaped pattern across layers, mirroring our correlation findings. Lower layers (0-3) produced poor place discrimination (63-64\%), performing worse than our MFCC-based model, while intermediate layers peaked at 80\% (layers 6-7) -- substantially improving over MFCC baselines though remaining below ground-truth toplines. Higher layers (9-12) show a marked decline in performance, with scores dropping to around 62\% at layer 12, below even the MFCC-based model.

Overall, this suggests that our model successfully learns articulatory trajectories that reflect the natural organization of consonants by place of articulation. Notably, intermediate wav2vec 2.0 layers (6-7) consistently yield the best performance across both evaluation metrics: they produce articulatory trajectories that both correlate most strongly with human trajectories and most effectively discriminate between different places of articulation.

\subsection{Multi-speaker setting}

Infants face a more challenging scenario than imitating their own voice: imitating diverse speakers with varied vocal tracts. We now investigate this more realistic multi-speaker setting.

\subsubsection{Organization along the place of articulation (multi-speaker)}

\begin{figure}[htbp]
    \centering
    \includegraphics[width=\columnwidth]{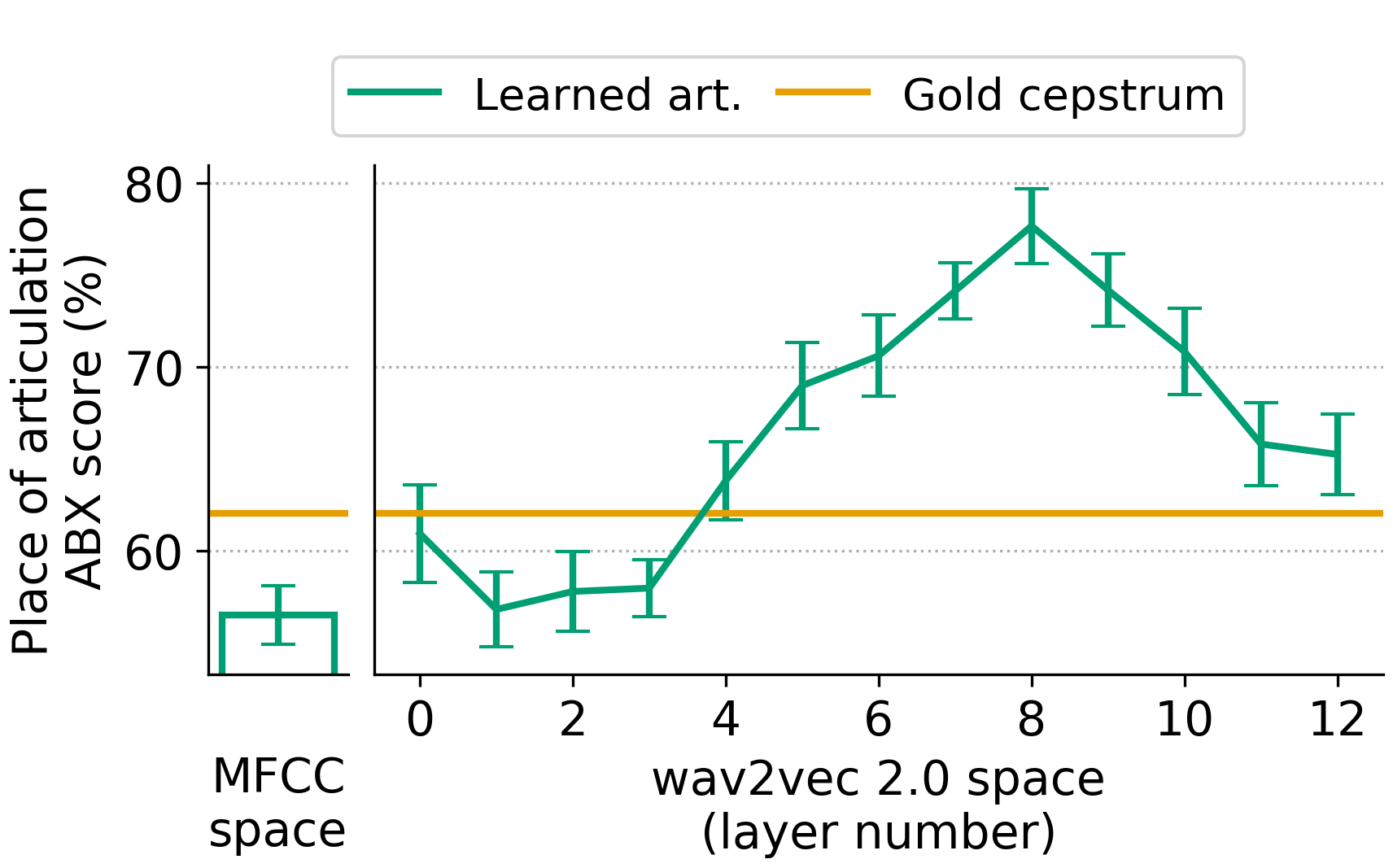}
    \caption{Place of articulation ABX scores (\%) obtained by the articulatory parameters learned by the model imitating in the MFCC feature space (left) versus the wav2vec 2.0 space (right). The horizontal orange line shows scores obtained by MFCCs extracted directly from the input audio. Error bars represent standard deviation across the 8 speakers of the Audiocite test set. All models are trained on the Audiocite training set.}
    \label{fig:8_speakers_6000_mn}
\end{figure}

Figure \ref{fig:8_speakers_6000_mn} shows place of articulation ABX scores for our multi-speaker imitation model. In the MFCC feature space, the model learns articulatory trajectories with a place discrimination score of approximately 56\%, below the acoustic baseline (orange horizontal line) of about 62\% obtained from using MFCCs directly for discrimination. This contrasts with our single-speaker results, where MFCC-derived trajectories outperformed the acoustic baseline. This suggests MFCCs contain sufficient information for articulatory learning within a single speaker but lack the speaker-invariant properties necessary for cross-speaker generalization.

In the wav2vec 2.0 feature space, we again observe a clear inverted U-shaped pattern across layers. Lower layers (0-3) produced articulatory trajectories with relatively poor place discrimination (56-58\%), performing similarly to the MFCC-based model. However, performance improved dramatically for intermediate layers, peaking at layer 8 with scores of approximately 77\% -- a substantial improvement over both the MFCC-based model and the acoustic discrimination baseline. Higher layers (9-12) show a decline in place discrimination performance, with scores dropping to around 65\% at layer 12, though still better than the MFCC-based model. The optimal layer shifts from layer 7 (single-speaker) to 8 (multi-speaker), suggesting higher-level representations benefit cross-speaker generalization by offering increased speaker invariance while maintaining sufficient phonetic detail. 

These results suggest that our model fails to learn human-like articulatory patterns from diverse speakers when representing speech using low-level acoustic features. However, intermediate layers of self-supervised models provide representations that enable effective learning of place-distinctive articulatory patterns across multiple speakers.

\subsubsection{What speech representations best support articulatory learning?}

So far, our results show that articulatory learning performance varies dramatically depending on which representations are used to encode speech input and optimize the objective. This raises an important question: What properties of speech representations enable successful articulatory learning?

\begin{figure}[h!]
    \centering
    \includegraphics[width=.95\columnwidth]{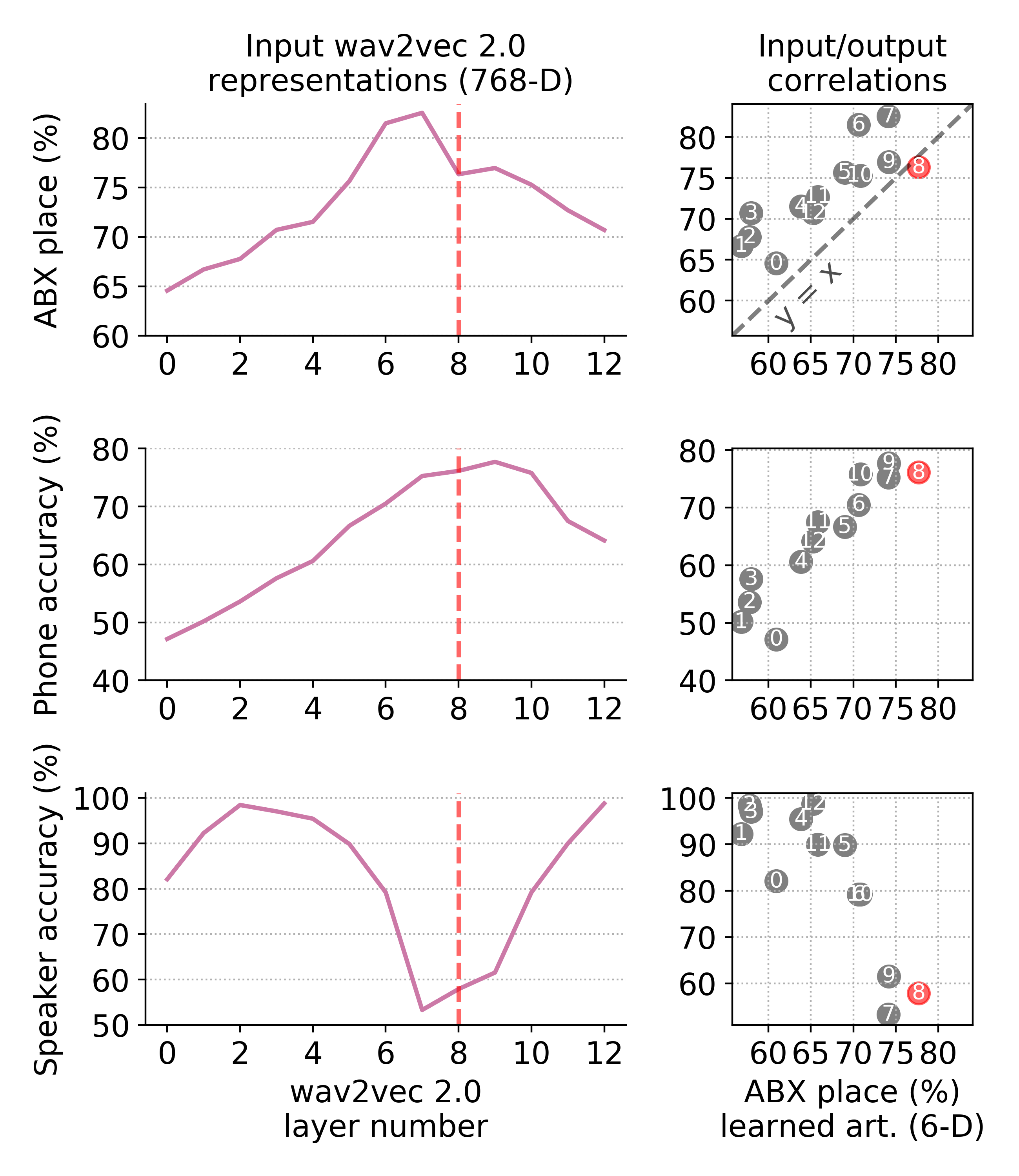}
        \caption{Left column: performance across wav2vec 2.0 layers (0-12) for place of articulation ABX (top), phone accuracy (middle), and speaker accuracy (bottom). The red dashed line indicates layer 8, which yielded articulatory parameters most discriminative of place of articulation. Right column: Correlation between each metric from the left panels and the place of articulation ABX scores obtained by the articulatory parameters learned by the model imitating in the wav2vec 2.0 space. Layer 8 is highlighted in red. ABX scores are computed on Audiocite \cite{felice2024audiocite} and phone and speaker identification accuracies are computed via linear probing on LibriSpeech \cite{panayotov2015librispeech}. All imitation models are trained on Audiocite.
    }
    \label{fig:representational_space}
\end{figure}

To answer this question, we conducted probing experiments to examine the information encoded at different layers of the wav2vec 2.0 model ($z$ depicted in Figure \ref{fig:model}). In the left column of Figure \ref{fig:representational_space}, we compute for each layer: 1) place of articulation ABX score: how well the representations discriminate between different places of articulation, 2) phone accuracy: how accurately phonetic categories can be recognized, and 3) speaker accuracy: how well speakers can be identified. Results are shown in Figure \ref{fig:representational_space}. The right column correlates these scores with the ABX discrimination performance of the learned articulatory space $\hat{a}$.

Layer-wise analysis reveals that place and phone accuracy peak at intermediate layers (6-9), while speaker accuracy shows the opposite pattern with minimum around layers 7-8. How are these linked to articulatory learning outcomes?

Correlation plots (right column) indicate that both place discriminability and phone accuracy positively correlate with articulatory learning performance. On the contrary, speaker accuracy shows a negative correlation -- layers with lower speaker discriminability correspond with better articulatory learning outcomes. Layer 8 (highlighted in red) appears to represent an optimal balance point where phonetic information is well-preserved while speaker-specific characteristics are minimized.

This suggests successful articulatory learning depends on representations that maintain strong phonetic discriminability while abstracting away speaker-specific information - precisely the properties exhibited by intermediate wav2vec 2.0. layers. 

\subsubsection{Intelligibility of the imitated speech}

Having analyzed articulatory trajectories, we now evaluate the intelligibility of the imitated speech.

\begin{figure}[htbp]
    \centering
    \includegraphics[width=\columnwidth]{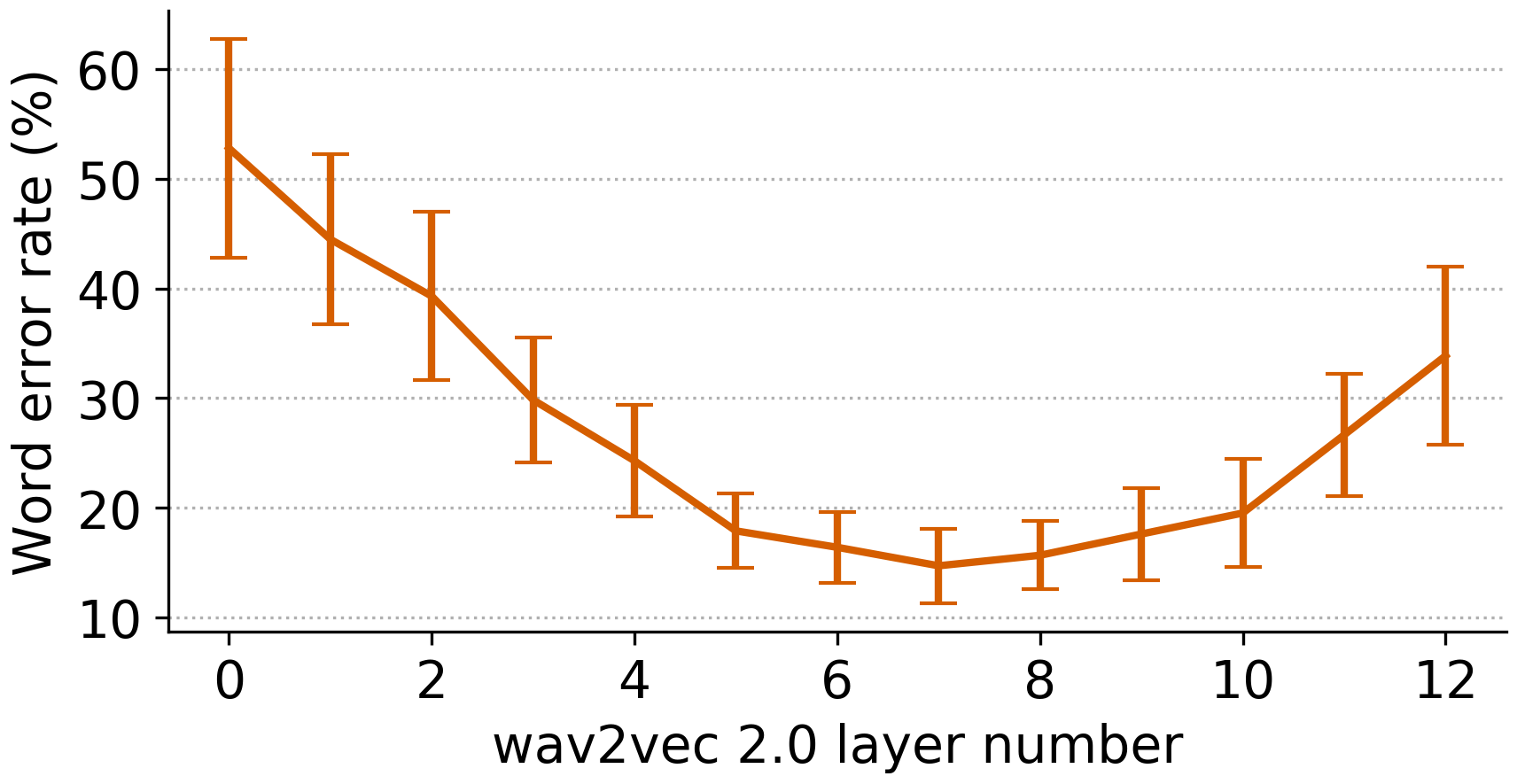}
    \caption{Word Error Rate (\%) of the imitated speech transcribed using Whisper-small when using different layers of wav2vec 2.0 as input speech representations. Error bars represent standard deviation across the 8 speakers from the Audiocite test set.}
    \label{fig:wer_imitated_speech.png}
\end{figure}

Figure \ref{fig:wer_imitated_speech.png} shows WER for speech imitated using different wav2vec 2.0 layers. The results reveal a clear U-shaped pattern: lower layers (0-2) yield high WER (40-53\%), intermediate layers perform best with WER reaching ~15\% at layer 7, and higher layers (10-12) show degraded performance (25-35\% WER). This pattern confirms our earlier findings: lower layers preserve acoustic detail but retain speaker-specific characteristics, causing high WERs. Intermediate layers (6-8) provide optimal representations that support cross-speaker articulatory mapping while abstracting away speaker characteristics. Higher layers lose phonetic discrimination, making them less effective.

\section{Discussion}

In this paper, we presented a computational approach to understanding speech imitation and production learning through a self-supervised model. Our model consists of a feature extractor that transforms speech into latent representations, an inverse model that maps these representations to articulatory parameters, and a pre-trained (and frozen) articulatory synthesizer and vocoder that generate speech from these parameters. The inverse model is trained to minimize the cosine distance between input and imitated speech representations, bridging the gap between perception and production.

We found that representations that simultaneously encode phonetic information while minimizing speaker-specific characteristics lead to a low-dimensional and interpretable articulatory space that effectively discriminates between different places of articulation. This, in turn, produces more intelligible imitated speech. Our findings align with \citet{cho2023evidence}, who found a high correlation between self-supervised speech representations and articulatory trajectories using supervised linear probing. Similarly, \citet{cho2024coding} demonstrated high-quality articulatory synthesis and cross-speaker generalization using self-supervised representations, but trained their inverse model with supervised learning on articulatory data. In contrast, our approach achieves similar physiologically aligned representations through entirely unsupervised learning without requiring any articulatory training data (the latter being used only for pre-training the synthesizer).


Our work provides an interesting perspective on research in the development of speech perception and production. According to \citet{kuhl2008phonetic}'s Native Language Magnet theory: ``sensory learning occurs first, based on experience with language, and this guides the development of motor patterns``. Our results are consistent with this view, as they demonstrate that representations that effectively discriminate between phonetic categories while being speaker invariant provide optimal foundations for articulatory learning. If this developmental sequence holds, then the increasingly specialized perception for native language phonetic contrasts that emerges between 6-12 months \cite{werker1984cross,kuhl2006infants} should provide the foundation for subsequent advances in speech production. This prediction is supported by \citet{hochmann2014invariance}, who used pupillometry to demonstrate that 6-month-olds, but not 3-month-olds, can solve the acoustic invariance problem by recognizing the same consonant across different vowel contexts. Crucially, this perceptual ability emerged before canonical babbling: only 3 of 14 six-month-olds who showed the perceptual effect had acquired canonical babbling, while several pre-babblers showed even stronger effects. Just as our model requires phonetically organized, speaker-invariant representations to learn effective articulatory mappings, infants may need to develop similar perceptual representations before their vocalizations can take on the systematic properties of their ambient language.

\section{Conclusion}

Our experiments demonstrate that articulatory imitation performance in our model is strongly influenced by the nature of input representations. Representations that effectively encode phonetic distinctions while minimizing speaker variability yielded substantially better articulatory trajectories and speech outputs. The superior performance observed with these balanced representations provides computational evidence supporting developmental theories proposing that perceptual learning of phonetic categories guides articulatory development. These findings suggest a possible mechanism through which infants might solve the acoustic-to-articulatory mapping problem: by developing representations that normalize across speakers while preserving phonetic contrasts, and then using these as targets for articulatory learning. Our approach could be extended by incorporating anatomical development of the vocal tract, intrinsically motivated exploration mechanisms, and more naturalistic training data. By continuing to refine computational models that bridge perception and production, we can gain a deeper understanding of the remarkable emergence of speech in early human development.

\clearpage
\section{Limitations}
\label{sec:limitations}

While our model captures important aspects of speech motor control development, several limitations should be acknowledged. Our computational findings demonstrate plausible mechanisms consistent with developmental theories; however, we cannot establish whether this reflects the temporal sequence of human development. Here, we discuss several limitations that need to be overcome to develop more developmentally realistic models.

First, our approach keeps the vocal tract parameters fixed, whereas infant vocal tracts undergo significant anatomical changes during development. Future work could incorporate a dynamically changing articulatory synthesizer to model how anatomical development interacts with speech production learning. Indeed, a more realistic model could either: (1) use a neural-network synthesizer with parameterized development stages, though this would require longitudinal articulatory data from infants, which remain particularly challenging to collect, or (2) rely on physical models of the vocal tract based on fluid dynamics and biomechanics that directly simulate how changing anatomical properties affect acoustic output \cite{birkholz2020printable,serrurier2023morphological}. While physical models would provide more interpretable parameters for developmental changes, they typically lack differentiability -- but see \citet{georges2024decode} for using a differentiable forward model that approximates the potentially non-differentiable synthesizer.

Second, our model does not incorporate vocal exploration or babbling phases that characterize infant speech development. Research shows that infant vocal exploration serves crucial purposes in establishing foundations for speech by forming new vocal categories \cite{yoo2024infant}. \citet{warlaumont2016learning}'s computational models demonstrate how infants might discover new speech sounds through self-guided exploration, with sounds that are more acoustically interesting receiving internal rewards that encourage further practice. This approach complements \citet{moulin2014self}'s work on curiosity-driven learning, where exploration is guided by seeking maximal information gain in the sensorimotor space. Future work could integrate these intrinsically motivated exploration mechanisms with our imitation framework to model the full developmental progression from early vocal play to intentional imitation.

Third, while our study made progress by training on multiple unknown speakers, future work should incorporate more naturalistic training data, as argued in \cite{lavechin2022reverse}. This presents challenges, including background noise, overlapping speech, and variable acoustic environments that better reflect what infants actually experience \cite{lavechin2024modeling}. On such noisy data, visual information could serve as a crucial guiding mechanism, providing stable anchor points when acoustic signals are ambiguous or degraded \cite{10.1162/neco_a_01264}. The visibility of articulatory gestures, particularly lip movements, offers consistent cues that could help constrain the mapping between acoustic input and articulatory configurations.

\section*{Acknowledgments} This work was supported by the France 2030 program (MIAI institute with projects ANR-23-IACL-0006 and ANR-19-P3IA-0003), the Simons Foundation International (034070-00033) and the COST Action @ University of East London (CA22111).

\bibliography{custom}
\clearpage  
\onecolumn
\appendix

\section{Examples of inferred articulatory trajectories}
\label{sec:appendixA}
\FloatBarrier
\vfill
\noindent\begin{minipage}{\textwidth}
    \centering
    \includegraphics[width=.8\textwidth]{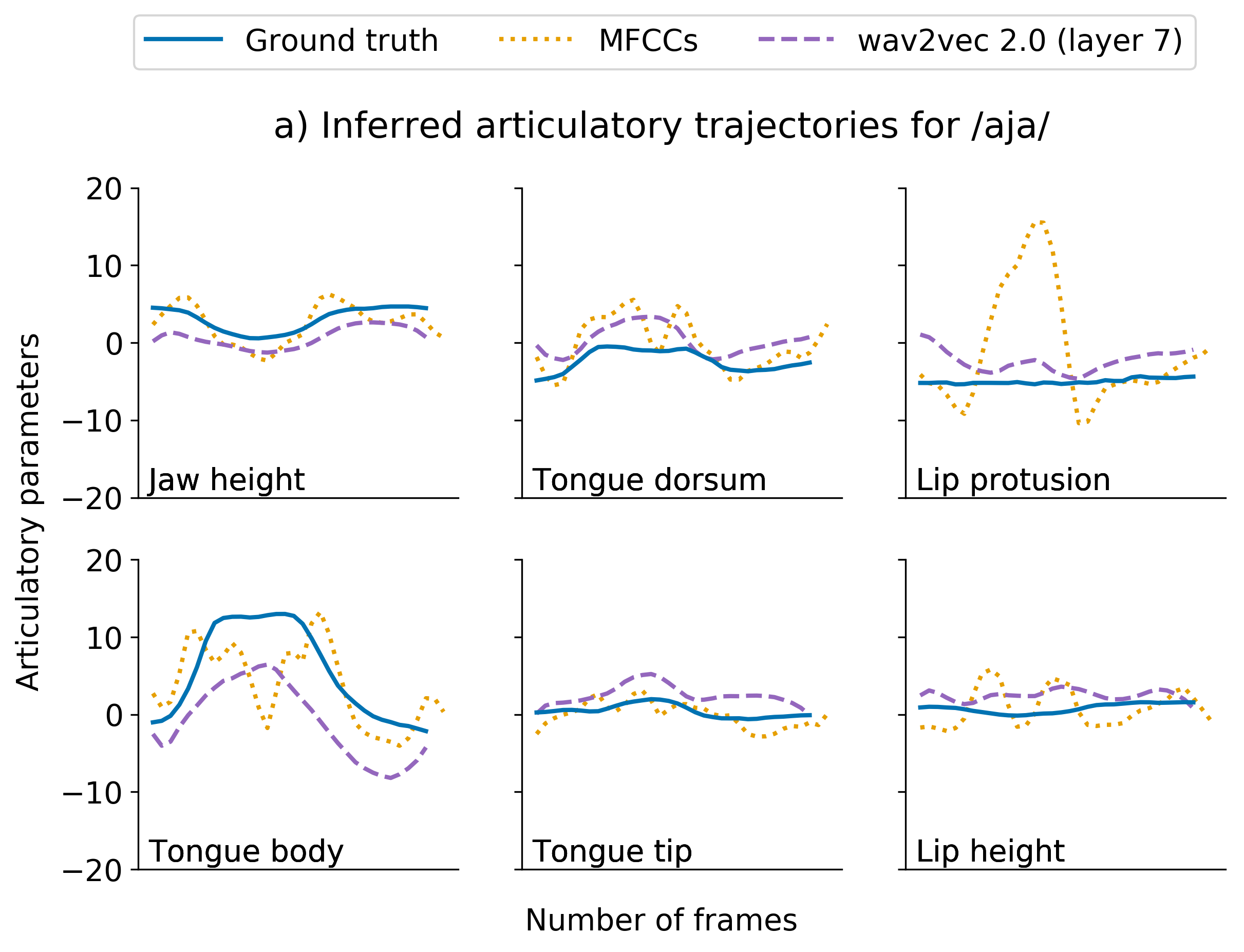}
    \hspace{1cm}
    \includegraphics[trim=0cm 0cm 0cm 1.5cm, clip, width=.8\textwidth]{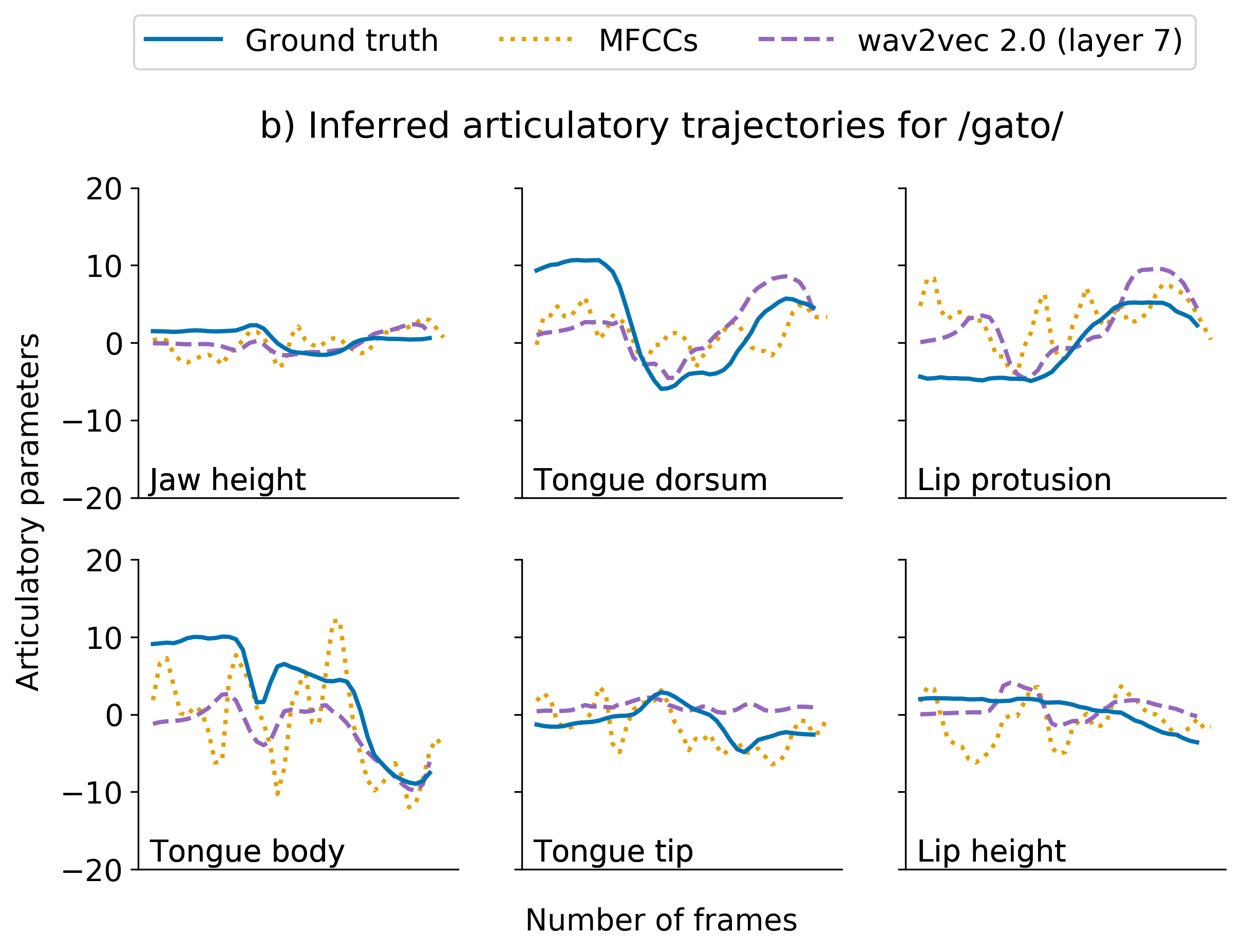}
    
    \captionof{figure}{Examples of articulatory trajectories inferred by the model imitating in the MFCC or the wav2vec 2.0 space compared to ground truth articulatory trajectories for a) /aja/ and b) /gato/.}
    \label{fig:aja_gato}
\end{minipage}
\vfill
\end{document}